# A new versatile in-process monitoring system for milling

http://dx.doi.org/10.1016/j.ijmachtools.2006.01.001

Mathieu Ritou*, Sebastien Garnier, Benoit Furet, Jean-Yves Hascoet

Institut de Recherche en Communications et Cybernetique de Nantes (IRCCyN) – UMR CNRS 6597

1 rue de la Noe, BP92101, 44321 Nantes Cedex 03, France

* Corresponding author. Tel.: +(33) 2 40 37 69 54, Fax: +(33) 2 40 37 69 30

E-mail address: Mathieu.Ritou@irccyn.ec-nantes.fr

**Abstract**

Tool Condition Monitoring (TCM) systems can improve productivity and ensure workpiece quality, yet, there is a lack of reliable TCM solutions for complex and flexible industrial manufacturing. TCM methods which include the characteristics of the cut seem to be particularly suitable for these demanding applications. In the first section of this paper, three process-based indicators have been retained from literature dealing with TCM. They are analyzed using a cutting force model and experiments are carried out in industrial conditions. Specific transient cuttings encountered during the machining of the test part reveal the indicators to be unreliable. Consequently, in the second section, intermittent monitoring is suggested. Based on experiments carried out under a range of different cutting conditions, an adequate indicator is proposed: the relative radial eccentricity of the cutters is estimated at each instant and characterizes the tool state. It is then compared with the previous tool state in order to detect cutter breakage or chipping. Lastly, the new approach is shown to be reliable when implemented during the machining of the test part.

# 1. Introduction

Machining problems, such as cutter breakage, excessive wear, chatter and collision, impede production consistency and quality. Loss can be significant, particularly when high added value parts like moulds and dies or aeronautical motor and structure parts are machined. They are manufactured in small batches or one-off productions. Thus, their machining should be monitored as soon as the first part is produced. Loss due to disturbance could be prevented, or at least limited, using an in-process Tool Condition Monitoring system (TCM). An accurate and reliable TCM system could increase savings of between 10% and 40% [1]. However, there is a lack of reliable TCM solutions for complex and flexible production in milling [2]; the subject of this paper. Part machining time can last for several days, without stopping during the off-duty hours of the operators. Thus, to prevent machining from being stopped too often, no false alarms can be allowed. The TCM system must therefore be completely reliable [3] as soon as the first part is machined.

**Fig. 1. TCM classification.**

Certain information is needed to evaluate process conditions. This is provided by one or several sensors which are placed in the machine tool. Various methods then allow analysis and decision-making, fig. 1. The teach-in method is used for mass production and most commercial TCM systems are based on this principle [4]. It requires the machining of a few parts (trial cuts) to measure a reference signal. Thresholds are then set on either side of the signal, based on heuristic knowledge [5]. As monitoring trial cuts is impossible, this is incompatible with flexible production [6].

It was suggested that the measured reference signal be replaced with an estimated one, using a cutting force model [7,8]. This enabled us to monitor the machining of the first part. The relevancy of this method relies on the accuracy of the force model. Till now, average cutting force values per spindle revolution were estimated and the gap between measured and estimated forces was significant. Furthermore, in milling, it is assumed that if a tooth is chipped or broken, this tooth removes a smaller volume of material than before breakage and the following one a larger volume [9]. Therefore, the cutting force per tooth period should be considered, rather than per spindle revolution. This method is consequently not suitable for cutter breakage detection.

The milling forces waveform has led various authors to feature extraction methods from the force signals of an incident. These methods are generic and applicable from the production of the first part. Many studies have been carried out on Artificial Intelligence (AI), e.g. neural networks, fuzzy logic. Neural networks or hybrid AI systems are viable for TCM [10]. Networks are trained using trial cuts. This then leads to the generalization problem: under other cutting conditions, the neural network may be unreliable [11]. Users may have to train the network again in order to monitor the machining of a new part [12].

Other authors suggested specific feature extraction of an incident using Digital Signal Processing methods, e.g. autoregressive filter [13,14], synchronized averaging [15,16], wavelet transform [17,18]. However, if basic process knowledge is ignored, it is harder to differentiate tool breakage from the effects of tool runout or transient cutting. Indeed, adequate force models have been developed [6] and geometric, kinematic, and mechanistic characteristics of the cutting process are, or could potentially be, controlled during milling operations [11]. They can be used to improve or simplify the signal-processing method and increase its reliability.

Little literature has been published on process-based signal processing, where characteristics of the cut were included (table 1). This method is particularly suitable for flexible production monitoring.

**Table 1. Summary of properties of process-based TCM criteria.**

At each spindle revolution, a force value is extracted for each tooth, before calculating the indicator value (fig. 2). Altintas and Yellowley [6] used the first and the second differences of mean forces, between adjacent teeth. It was shown that it was impossible to distinguish tool breakage from cutter runout [19]. Lee et al. [19] added a new indicator to the 1$^{st}$ order autoregressive filter proposed by Altintas [14]: the relative variation of average tooth force, between two consecutive revolutions. They introduced the idea that each tooth can be monitored individually. But the criteria [6,19] were affected by changes in cutting conditions and therefore tool breakage detection was unreliable [20]. Kim and Chu [20] proposed the Tool Failure Index (TFI), which is the ratio of peak-to-valley cutting forces between adjacent teeth, divided by its own past average ratio. The average ratio is intended to prevent the TFI from cutting conditions changes. In this paper, the TFI is retained for further experiments to examine whether it is unaffected by changes in cutting conditions. Lastly, Deyuan et al. [21] proposed two indicators: the peak rate $K_m$ is the ratio of the difference to the sum between peak forces of adjacent teeth. The relative eccentricity rate $B_m$ is similar to the ratio of tooth eccentricity to maximum chip thickness. The authors specified that the indicators were independent of cutting conditions. As for TFI, the two indicators have been retained.

Generally, few experiments are carried out to evaluate the relevancy of the criteria. Nevertheless, machining does not comply with High Speed Machining cutting conditions and trajectories used for complex and flexible industrial production. They consist of a simple straight path, conducted under an over-limited range of cutting conditions. The latter are generally low, e.g. cutting speeds of less than 40 m/min while machining carbon steel or aluminium alloy [8,15,17,18,19,22,23]. In this way, significant and sudden changes are encountered neither in cutting conditions, nor in cutting forces, respectively. Thus, there is a lower risk that cutting forces transients would be misinterpreted and generate false alarms. So, there is generally a lack of experiments under industrial cutting conditions and trajectories. This is also the case with the retained criteria.

In the first section of this paper, we will present a study of three process-based TCM indicators, extracted from literature. It is verified whether they are unaffected by changes in cutting conditions, so as to evaluate their relevancy for the monitoring of complex and flexible productions. Experiments were carried out under various real industrial machining settings. The criteria were found to be unreliable, due to misinterpretation of sudden changes in cutting conditions. Therefore, in the second section, intermittent monitoring is suggested, to tackle the problem of reliability. A new approach is proposed based on our experiments of milling forces under a range of cutting conditions. Relative radial eccentricity is estimated and this characterizes instant tool state. Unlike the criteria extracted from literature, this new approach was successfully implemented using the same experiments.

**Figure 2. Chip thickness and forces in milling.**

## 2. Criteria extracted from literature

*2.1. Definition* [20,21]

Peak $F_j$ and peak-to-valley $PV_j$ values are extracted from milling resultant forces, for each tooth $j$ and for each spindle revolution (fig. 2). $\overline{PV_j}$ represents the mean of $PV_j$ over the last ten spindle revolutions whereas $\overline{F}$ is the mean of the peak forces during the current spindle revolution. Z is the tooth number. Then, the criteria for each tooth and each spindle revolution are calculated as follows:

$$TFI_j = \frac{PV_j / PV_{j-1}}{\overline{PV_j / PV_{j-1}}} \qquad Km_j = \frac{F_j - F_{j-1}}{F_j + F_{j-1}} \qquad Bm_j = \sum_{k=2}^{Z}\left(\frac{F_k}{F} - 1\right) \qquad (1)$$

The TFI takes into account the last ten spindle turns to prevent it from tool runout and changes in cutting conditions. So, whether the tool is new or worn, TFI=1 during steady cuts. It focuses on sudden force transients to detect cutter breakage. After an event, it returns to 1. Thus, it is important to distinguish transient cut and problem correctly when forces vary. Bm and Km characterize the process at any given spindle revolution and their values are compared to fixed thresholds [21,24].

*2.2. Analytic study*

In order to determine what the criteria depend on, Sabberwal [25] force models are used, where $k_t$ and $k_r$ are constants, $h_c(\varphi) = f_z.\sin(\varphi)$ the instant chip thickness [26], $a_p$ the depth of cut, $f_z$ the feed per tooth. An overview and particulars of force models were published [27,28]. However, for an early study, Sabberwal force models were used.

$$\begin{cases} F_t = k_t.h_c(\varphi).a_p \\ F_r = k_r.F_t \end{cases} \qquad (2)$$

The term $\varepsilon_j$ defines tooth radial eccentricity, the influence of the cutter shape, tool and spindle runout and the amount of cutter chipping. Relative radial eccentricity $\Delta\varepsilon_j$ is introduced in the expression of instant chip thickness removed by tooth #j [21,29]:

$$h_{c\,j}(\varphi) = f_z.\sin\varphi + \varepsilon_j - \varepsilon_{j-1} = h_c(\varphi) + \Delta\varepsilon_j \qquad (3)$$

Using the hypothesis that only one tooth participates in the cut at the same time ($h_c$ is the maximum chip thickness and K a specific cutting coefficient),

$$PV_j = F_j = K.a_p.(h_c + \Delta e_j) \qquad (4)$$

An equivalent formulation of criteria is obtained, depending on cutting conditions:

$$TFI_j = \frac{h_c + \Delta e_j}{h_c + \Delta e_{j-1}} \bigg/ \overline{\frac{h_c + \Delta e_j}{h_c + \Delta e_{j-1}}} \qquad Km_j = \frac{\Delta e_j - \Delta e_{j-1}}{2.h_c + \Delta e_j + \Delta e_{j-1}} \qquad Bm_j = \frac{e_j}{h_c} \qquad (5)$$

In the case of breakage or chipping of teeth #$j$, $\varepsilon_j$ decreases. So, criteria formulation allows them to be detected. The influence of $h_c$ can be seen for each indicator. Therefore, if the feedrate or the width of cut varies during the machining of a part, the criteria should vary and this could affect their reliability.

This explains why such experiments have been carried out, where feedrate and width of cut vary during machining. A corresponding test part was designed. A pocketing operation was chosen with a zigzag strategy, allowing up milling and down milling. This comprises both simple and sharp turns (figure 3), where the feedrate should drop, due to the limited acceleration available on the machine axes [30]. A contouring path finishes the pocket.

**Figure 3. Test part and toolpath (on the left). Experimental set-up (on the right).**

*2.3. Experiments*

Cutting force signals were measured using a 9257A Kistler quartz three-component dynamometer, sampling at 64 kHz. The dynamometer was mounted between the workpiece and the table of a Sabre Cincinnati machining centre. The X and Y axes position encoders were measured using a sample frequency set at 500 Hz [31]. The workpiece was made of 7075 aluminium alloy. The parameters were the feed per tooth (0.08, 0.12, 0.16 and 0.2 mm/rev/tooth), the width of cut (15, 40, 65, 90% of tool diameter) and the tool (a 32 mm diameter with 2 inserts and a 20 mm diameter endmill with 3 flutes). Different sets of these parameters were tested at every level of the workpiece. Depth of cut was 2.5 mm. Since cutting speed was 650 m/min, spindle speeds were 6 500 and 10 000 RPM, and feedrates ranged from 1 to 6 m/min. So, unlike many studies, the experiments were carried out under real industrial cutting conditions.

The X and Y force components were low-filtered at twice the tooth passing frequency before calculating the resultant force. Then, force minimums and maximums were calculated for each tooth and for each spindle revolution, to evaluate $F_j$ and $PV_j$. Based on axes encoder measurements, the instant feedrate Vf was calculated as well as the instant width of cut $a_e$. The edges of the workpiece were discretized every 0.05 mm. Then, for each new tool position, intersections with the swept volume of the tool led to the entry and exit angles of the teeth and then the instant width of cut $a_e$ was obtained. The Z axis of figure 4 represents the instant cutting conditions (Vf and $a_e$) during the machining of the pocket. It reveals that sudden changes occur during turning.

**Figure 4. Instant feedrate and width of cut during pocket machining.**
**(CAM settings: Vf=3.6 m/min, $a_e$=65%, tool ø20mm)**

*2.4. Implementation of criteria*

The criteria were applied to the force signals measured during the machining of the pocket. The instant feedrate and instant width of cut allowed a better understanding of the behaviour of these criteria (cf two first graphs in figure 5). The 3$^{rd}$ graph represents the resultant cutting forces (blue curve). Peak to valley values were extracted from the latter for each spindle revolution and each tooth (red, green, and black curves). Then, the TFI was calculated from these values (4$^{th}$ graph). In this way, peak values per tooth were extracted (5$^{th}$ graph) and Deyuan et al.'s criteria were calculated.

**Figure 5. Behaviour of the criteria during rough milling of pockets.**
**(CAM settings: Vf=3.6 m/min, $a_e$=65%, tool ø20mm)**

During steady cuts, TFI = 1. Bm and Km take a set of values for each tooth. If a problem occurs, force signals are modified and the indicator values change, allowing incident detection [20,21]. Note that Bm always goes beyond the threshold proposed by Deyuan et al. during finish milling because the chip thickness is too inferior to the cutter eccentricity [24]. There would be a permanent false alarm, in this situation.

During simple turns (e.g., zone 1 fig. 3 or 1$^{st}$ turn fig. 5), the feedrate of the axes of the machine slows down whilst turning because the acceleration of each axis is limited [30]. Therefore, the feedrate decreases. It was found that $a_e$ also varies. Towards the end of a straight path preceding a turn, the width of cut increases due to the material left behind during the previous path, as between positions *a* and *b* in figure 6. Turning begins at position *b*. Since the acceleration available on machine tool axes is limited, feedrate slows down and the controller adds a portion of circle in the corner, to allow turning with a lower but acceptable feedrate [30]. Disregarding spindle rotation, the tool turns around an axis located on its left. Its right side moves into the material, increasing the exit angle f$_s$ and leading to down milling. On the left hand side, material has already been partially removed, so the entry angle f$_e$ decreases a little and $a_e$ reaches a peak at position *c*. During the second half of the turn, $a_e$ decreases because f$_e$ decreases further and f$_s$ has reached its maximum. After *d*, the link path begins, the tool penetrates the material and the width of cut can reach a full diameter immersion. Under these moderate changes in cutting conditions, criteria variations are negligible (figure 5) and tool breakage can be detected regardless [24].

**Figure 6. Variation of the instant width of cut during turns.**

On the contrary, during sharp turns (2$^{nd}$ turn in figure 5), the changes in cutting conditions are more significant. The drop of $a_e$ is substantial because, in the second half of the turn, most of the material on the left of the tool has already been removed.

However, the influence of $a_e$ is negligible: beyond $a_e = 50\%$ of tool diameter, $max\{\sin f\}=1$. So, maximum chip thickness ($h_c=f_z*max\{\sin f\}$) and peak forces are theoretically unaffected. In the $a_e$ graph in figure 5, a minimum of 35% is reached. This corresponds to f =75° and $max\{\sin f\}=0.95$, i.e. 95% of its previous value. As the $a_e$ minimum is reached when the feedrate returns to a medium value, the influence of $a_e$ is negligible in this case, unlike during entry and exit transients and finish paths.

During sharp turns, the fall in feedrate is significant and, due to the relative cutter eccentricity ? $e_j$, some of the teeth remove hardly any material. That only a few teeth of the tool participate in the cut is quite usual: this happens during entry and exit transients, sharp turns and finish paths (according to cutting conditions). Figure 5 reveals that, in these cases, false alarms would have been sounded. This can be explained by the equivalent formulation of the criteria (eq. 5) [24]. Consequently, the criteria are unreliable during significant changes in cutting conditions. This contradicts what the developers of the criteria [20,21] suggested.

*2.5. Conclusions of the studied criteria*

It was shown that the current process-based TCM criteria from literature are unreliable. Although during steady cut or moderate changes in cutting conditions, reliable tool breakage detection could be carried out, faulty detection would be encountered when some of the teeth do not participate in the cut, namely during entry and exit transients, sharp turns and finish paths (according to cutting conditions). The main problem of the TCM criteria is correctly distinguishing transient cuts and problems, when force transients occur.

For example, about 6 false alarms per pocket level were found applying the criteria to the force signals measured during our experiments, whereas no disturbances occurred during machining. The part is composed of 10 levels. Therefore, 60 false alarms would result during the complete machining of this relatively simple part. Numerous false alarms would be sounded during the machining of a complex industrial part. The TCM system would soon be switched off [4]. This is the reason why a new approach is suggested in the following section: to solve the problem of reliability.

**Figure 7. False alarms, applying criteria [20,21] to the machining of a pocket level.**

## 3. Intermittent monitoring of cutter eccentricity

In this section, the authors will suggest intermittent monitoring of the tool state. The principle and restrictions of the TCM method are specified. Then, a suitable indicator which estimates the relative cutter eccentricity of the teeth is developed based on a milling forces study.

*3.1. Principle*

Numerous reasons may cause transients in milling forces. Some need the reaction of a TCM system, e.g. cutter breakage or chipping, excessive wear, chatter or collision, whereas others do not, e.g. entry or exit transients, turns, steps, hard points in the workpiece material, chip congestion and recycling, dynamic phenomena of the cut or cutter micro-welding. Interpreting every case from force signals is arduous and somewhat unreliable. This is why intermittent monitoring is proposed in this paper: only the zones where the TCM system can perform reliably are monitored.

The tool state is characterized by a set of parameters. As long as no problem occurs, the estimated tool state is assumed to remain identical (whatever the cutting conditions). Monitoring is paused during estimated tool state variation and resumes when stability returns. If the tool state is different, a problem has occurred. In this way, decisions are made with certainty. The authors suggest that long-term effects such as tool wear could also be detected, by periodically comparing the current tool state with the initial tool state.

A suitable indicator for intermittent monitoring is required. Whether the tool is new or damaged, TFI=1 during steady cuts and Deyuan et al.'s criteria vary by 140% under the range of cutting conditions of our experiments presented in §3.2; so, these criteria are incompatible with intermittent monitoring. Therefore, it is suggested that the relative radial eccentricity of cutters ? $e_j$ characterizes tool state, enabling detection of cutter breakage, chipping, and potentially reverse displacement of the cutting edge due to wear. Tool state is estimated using cutting force signals. As noted earlier, an average of forces during one spindle revolution is inaccurate for cutter breakage detection. Instant forces could be affected by industrial conditions and more difficult to implement. So, one value per tooth and per spindle revolution seems suitable. If $F_j$ is considered, rather than $PV_j$, it can be seen that the gap between the curves for each tooth is more consistent (cf. figure 5). Therefore characterizing tool state using force peaks is more relevant.

*3.2. Force model*

Since the estimated tool state has to be identical whatever the cutting conditions, a study of peak forces was carried out so as to develop an indicator independent of cutting conditions. A full factorial design of experiments was made, with 2 factors at 4 and 8 levels (respectively $f_z$ 0.08, 0.12, 0.16, 0.2 mm/rev/tooth; and $a_e$ 100%, up milling 15, 40, 65% and down milling 15, 40, 65, 90 % of tool diameter). The tool with 2 inserts was used and other experimental parameters can be found in §2.3. For each steady cut corresponding to a 2-element set ($f_z$,$a_e$), cutting force peaks $F_j$ were extracted for each tooth $j$, fig. 8.

**Figure 8. Cutting force peaks for each tooth, under various feedrates and widths of cut (on the left). Linear least square fitting of force peaks per tooth, under fixed width of cut (on the right).**

For a given $a_e$, it was found that a constant slope $k_{cj}$ could be identified for each tooth (fig. 8). Considering eq. 6, $k_{cj}$ and $b_{ij}$ were obtained by linear least square fitting, with correlation coefficients of at least $R^2$=99.3% and 11% of deviation of $k_{cj}$, disregarding $a_e$=15% which is close to the limits of the model. Although $b_{ij}$ varies, a permanent gap between force peaks of tooth #1 and #2, was observed. This is due to the radial eccentricity of the teeth, which distributes the material to be removed unevenly between teeth.

$$F_j = k_{c\,j}.h_c + b_{ij} \qquad (6)$$

*3.3. Indicator formulation*

As an invariant $k_c$ can be identified for each tooth when a given workpiece material is machined under various cutting conditions, the link between radial eccentricity and cutting forces can be summarized as in figure 9.

**Figure 9. Link between relative radial eccentricity and cutting forces, locally.**

The force reference $F^*$ corresponds to the mean of cutting force peaks (during the current spindle revolution) if every tooth removed the same volume of material, i.e. if there was no radial eccentricity. Eq. 3 implies that the larger the radial eccentricity is, the larger the chip thickness and the larger the force peak. They are linked by

$$\Delta F_j = F_j - F^* = k_{c\,j}.\Delta e_j \,, \text{ with } \sum_{j=1}^{Z} \Delta e_j = 0 \qquad (7)$$

Then, current ∆$e_j$ can be estimated from force peaks of the current spindle revolution, using eq. 8. In order to respect eq. 7, F* is defined as follows:

$$\Delta e_j = \frac{F_j - F^*}{k_{cj}}, \text{ where } F^* = \frac{\sum_{j=1}^{Z} F_j / k_{cj}}{\sum_{j=1}^{Z} 1/k_{cj}} \quad (8)$$

Figure 10 shows that the relative radial eccentricity of tooth #1 is estimated at 35 ±5 µm, under a wide range of cutting conditions (feed varying from 0.08 to 0.2 mm/rev/tooth and width of cut from 15 to 100% of tool diameter, both up milling and down milling).

**Figure 10. Estimated relative radial eccentricity under various cutting conditions.**

## 4. Implementation using the test part experiments

### 4.1. Initial tool state

The first step of the proposed method consists in identifying the $k_{cj}$ constants, which are specific to the tool and the material of the workpiece. A few straight paths are machined on a small part made of the same material as the workpiece, over a variety of feeds and widths of cut. Then, force peaks per tooth are extracted and $k_{cj}$ is determined by linear least square fitting. The initial tool state is also estimated.

**Figure 11. Estimated relative radial eccentricity while implementing the intermittent monitoring method with pocket machining. (CAM settings: Vf=2.08 m/min, $a_e$=65%, tool ø32mm)**

### 4.2. Monitoring during test part machining

Once $k_{cj}$ is identified, the current ∆$e_j$ is estimated from force peaks and eq. 8, during the machining of any part, enabling TCM. The monitoring is paused during transients of the estimated ∆$e_j$. Transients are relatively short; they last about 0.1 s. In figure 11, 90% of machining time is monitored (95%, disregarding entry and exit transients) and a rapid reaction is possible in the case of problem. Note that the estimated ∆$e_j$ remains constant during most width of cut variations, fig. 11.

Even if the TCM is sometimes interrupted for 0.1 s (e.g. during toolpath turnings), the proposed method allows reliable monitoring of any flexible production, preventing false alarms.

## 5. Conclusions

From the TCM literature, three process-based criteria were retained for further experiments, using industrial cutting conditions and trajectories. It was found that they were unreliable in several cases, i.e. when some of the teeth do not participate in the cut, as during entry and exit transients or sharp turns. Thus, they are affected by changes in cutting conditions. Calculated using cutting force signals, the criteria fail to distinguish correctly between problems and particular transient cuts. So, until now, there were no efficient and reliable solutions for the monitoring of flexible production in milling.

Since interpreting certain transient cuttings correctly is arduous, the authors suggested intermittent monitoring, where the TCM would be suspended during force transients, ensuring reliability. The tool state is calculated using force peaks per tooth during every spindle revolution and is characterized by the relative eccentricity of the teeth. The tool state is memorized and compared with the previous one. The method implies that the estimation has to be independent of cutting conditions. This is the reason why we have proposed a new indicator, presented in this paper and based on an experimental study of the milling forces under a wide range of cutting conditions. The method was then successfully implemented using the same experiments as for the aforementioned criteria, i.e. under industrial cutting conditions and trajectories. It was shown that, using only force signals, intermittent but reliable monitoring of flexible production is possible.

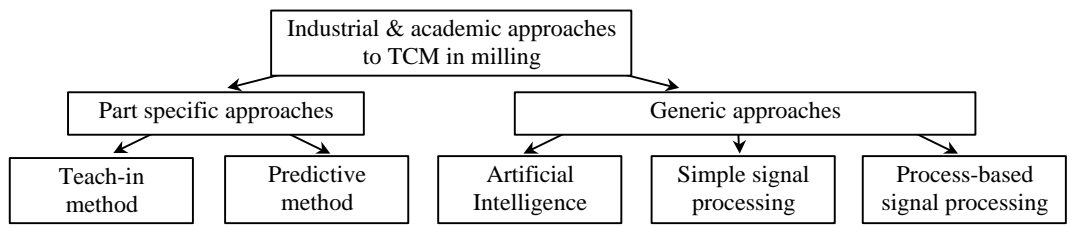

**Fig. 1. TCM classification.**

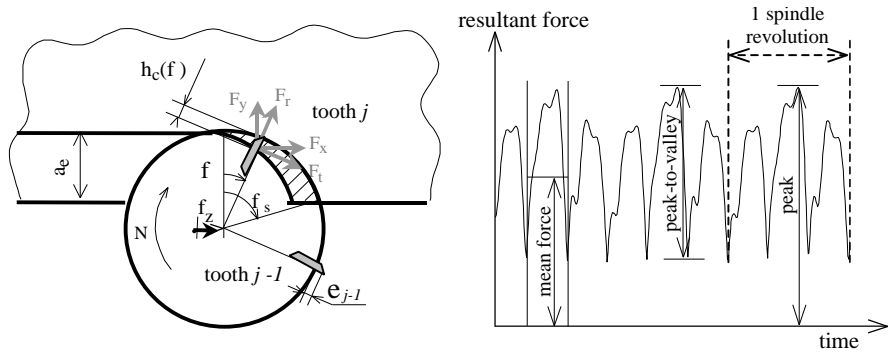

**Figure 2. Chip thickness and forces in milling.**

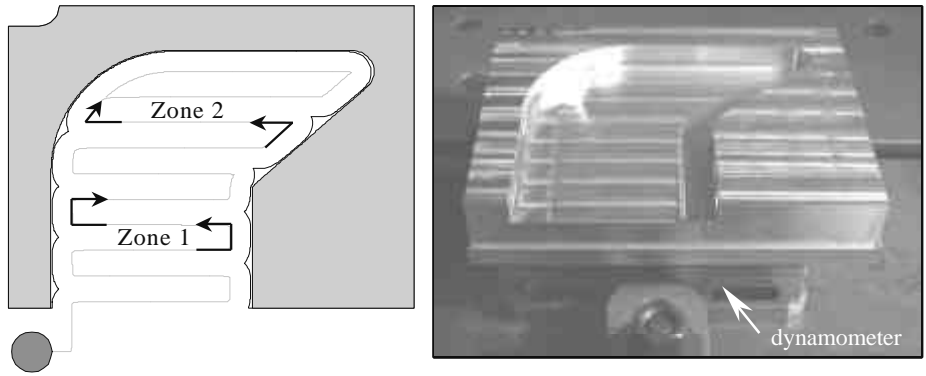

**Figure 3. Test part and toolpath (on the left). Experimental set-up (on the right).**

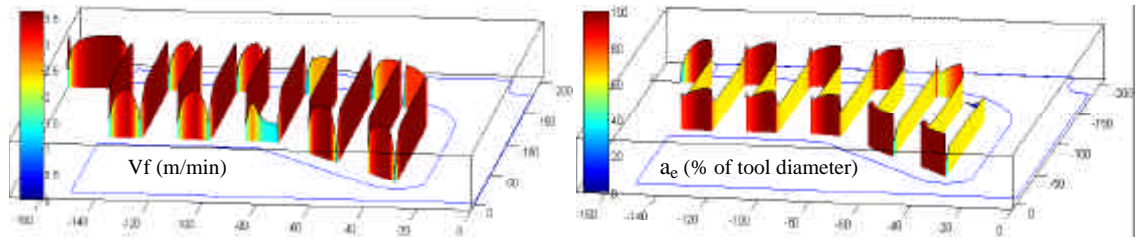

**Figure 4. Instant feedrate and width of cut during pocket machining.
(CAM settings: Vf=3.6 m/min, ae=65%, tool ø20mm)**

(this figure is intended to be reproduced in black-and-white)

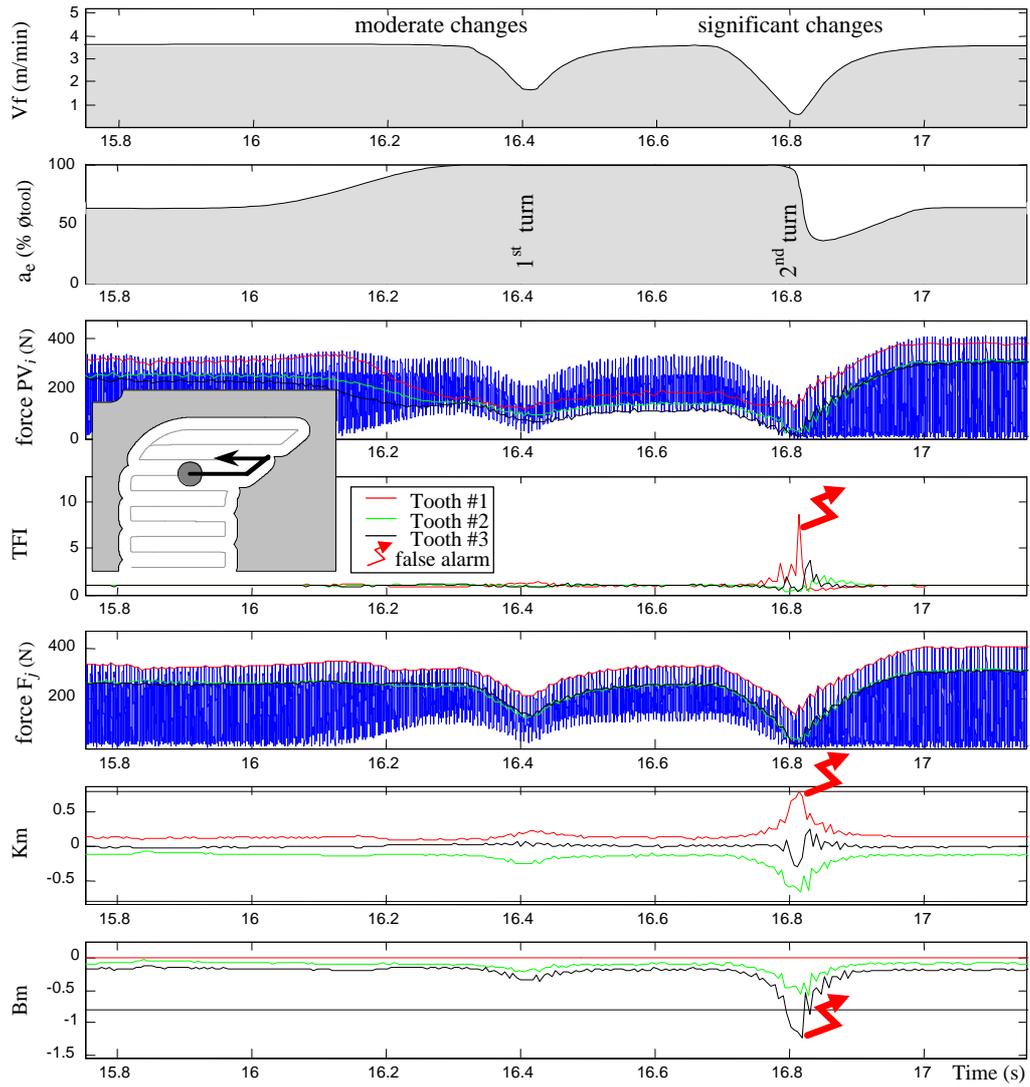

**Figure 5. Behaviour of the criteria during rough milling of pockets.**
**(CAM settings: Vf=3.6 m/min, $a_e$=65%, tool ø20mm)**

(this figure is intended to be reproduced in black-and-white)

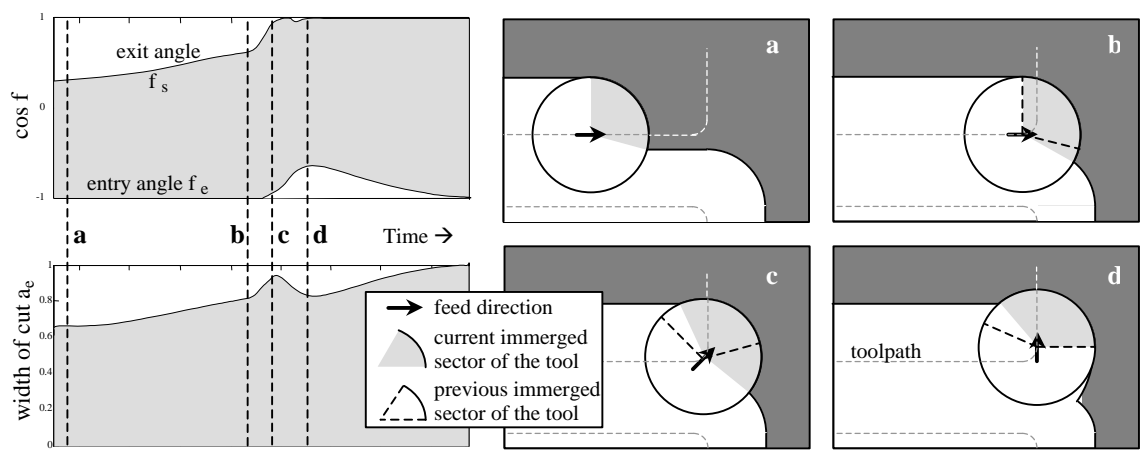

**Figure 6. Variation of the instant width of cut during turns.**

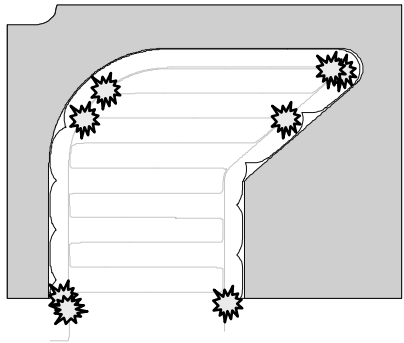

**Figure 7. False alarms, applying criteria [20,21] to the machining of a pocket level.**

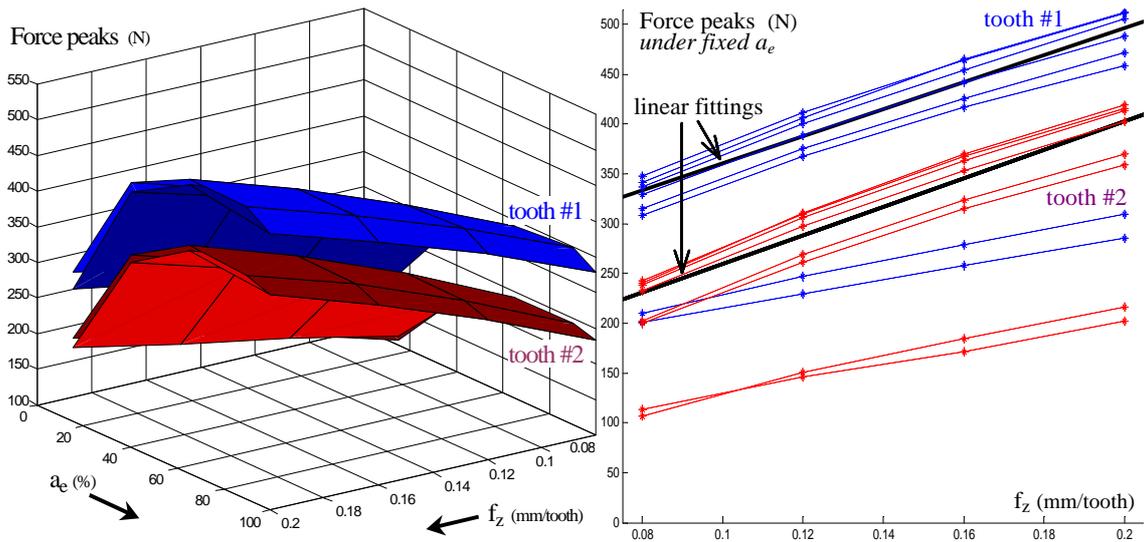

**Figure 8. Cutting force peaks for each tooth, under various feedrates and widths of cut (on the left). Linear least square fitting of force peaks per tooth, under fixed width of cut (on the right).**

(this figure is intended to be reproduced in black-and-white)

**Figure 9. Link between relative radial eccentricity and cutting forces, locally.**

(this figure is intended to be reproduced in black-and-white)

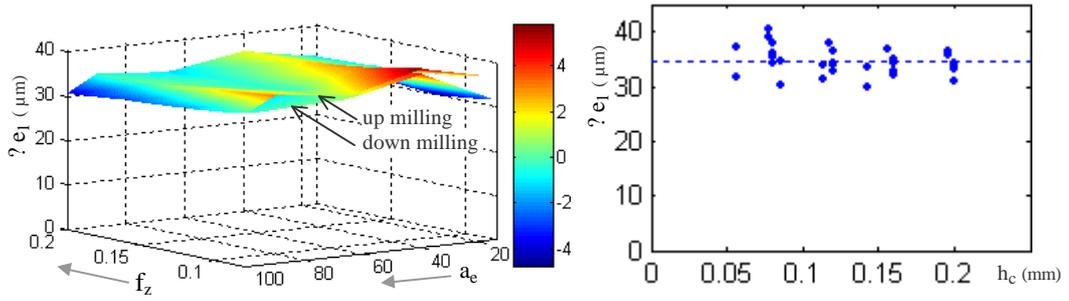

**Figure 10. Estimated relative radial eccentricity under various cutting conditions.**

(this figure is intended to be reproduced in black-and-white)

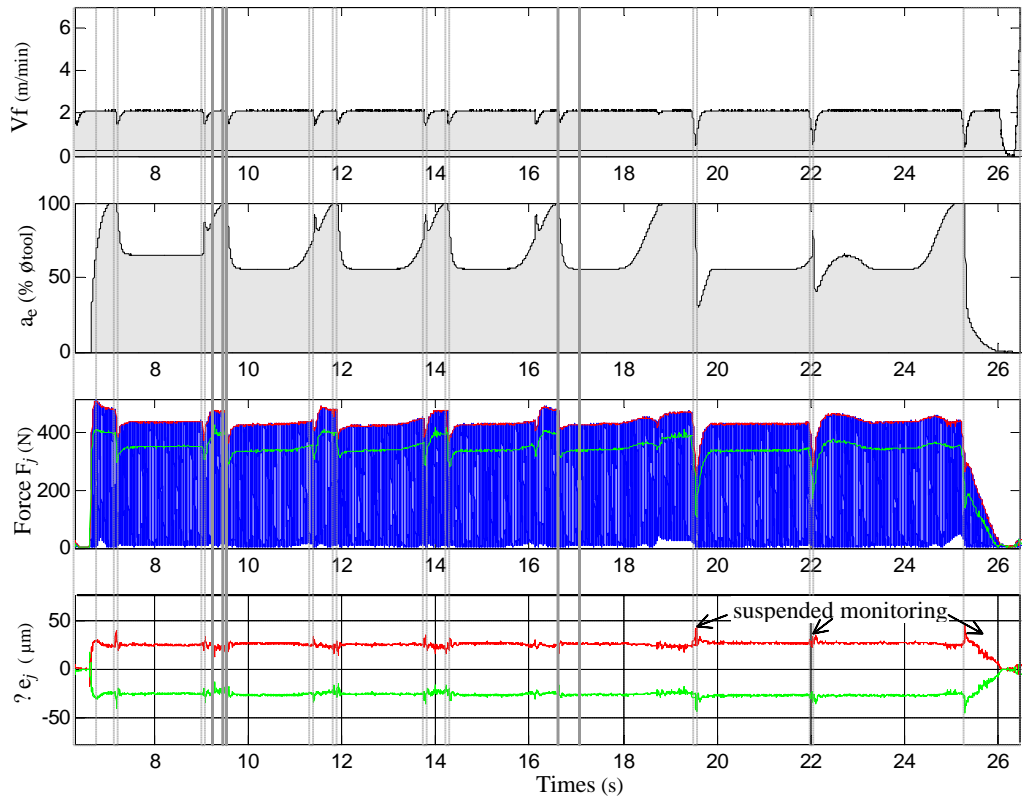

**Figure 11. Estimated relative radial eccentricity while implementing the intermittent monitoring method with pocket machining. (CAM settings: Vf=2.08 m/min, $a_e$=65%, tool ø32mm)**

(this figure is intended to be reproduced in black-and-white)

| Process-based TCM criteria | Cutter breakage or chipping detection | Unaffected by cutter eccentricity | Unaffected by changes in cutting conditions |
|---|---|---|---|
| Altintas Yellowley [6] | ✓ | X | X |
| Lee et al. [19] | ✓ | ✓ | X |
| Kim Chu [20] | ✓ | ✓ | ?? |
| Deyuan et al. [21] | ✓ | ✓ | ?? |

**Table 1. Summary of properties of process-based TCM criteria.**